%% file: 0-paper.tex
\documentclass{sigchi}
\usepackage{balance, graphics} 
\usepackage[T1]{fontenc}
\usepackage{txfonts, mathptmx}
\usepackage[pdftex]{hyperref}
\usepackage{color, booktabs, textcomp}
\usepackage{microtype, ccicons, tabularx}
\usepackage{todonotes, enumerate}
\usepackage{csquotes}
\usepackage{rotating, datetime}
\usepackage{placeins}
\usepackage{enumitem}
\usepackage{calc}

\usepackage{multirow, array, rotating}

\input{z-commands}

\def\plaintitle{How the Experts Do It: Assessing and Explaining Agent Behaviors in Real-Time Strategy Games}

\def\emptyauthor{}

\def\plainkeywords{
Explainable AI;
Intelligent Agents; 
RTS Games; 
StarCraft; 
Information Foraging
}

\definecolor{linkColor}{RGB}{6,125,233}
\hypersetup{%
  pdftitle={\plaintitle},
  pdfauthor={\emptyauthor},
  pdfkeywords={\plainkeywords},
  bookmarksnumbered,
  pdfstartview={FitH},
  colorlinks,
  citecolor=black,
  filecolor=black,
  linkcolor=black,
  urlcolor=linkColor,
  breaklinks=true,%
}

\begin{document}

\title{\plaintitle}


\numberofauthors{1}
\author{%
  \alignauthor{Jonathan Dodge, Sean Penney, Claudia Hilderbrand, Andrew Anderson, Margaret Burnett}\\
    \affaddr{Oregon State University}\\
    \affaddr{Corvallis, OR; USA}\\
    \email{ $\lbrace$ dodgej, penneys, minic, anderan2, burnett $\rbrace$@oregonstate.edu }\\
}

\maketitle{}

\FIXME{Draft status\\
\begin{tabular}{l|p{.4\linewidth}|l|l|l}
				& Page budget & Draft\\\hline
Abstract		& ok & 2.5 \\
Introduction 	& ok (to bot p2c1 including Fig1) & 2.5 \\
Background		& ok (3/4-1 page) & 2.4 \\
Methodology		& ok (1 pg including Table 1) & 2.5 \\
Results:Patches	& ok (1.75pg including Tab2+Fig2) & 2.5 \\
Results:Nav		& ok (1.5pg including Fig 3-5) & 2.2 \\
Results:LimDey	& probably ok & 2.2 \\
Results:Content & get down to 1.5-1.75pg & ? \\
Conclusion		& ok (about 1/4 page) & 2.5	\\
\end{tabular}
\{
}

\FIXME{task list\\
High level goals: paper is too starcraftish, If you are building an Ai shoutcaster, it should do what ours did
\begin{enumerate}

\item FIX THE MACROS, need some \%
\end{enumerate}
}

\begin{abstract}
How should an AI-based explanation system explain an agent's complex behavior to ordinary end users who have no background in AI?
Answering this question is an active research area, for if an AI-based explanation system could effectively explain intelligent agents' behavior, it could enable the end users to understand, assess, and appropriately trust (or distrust) the agents attempting to help them.
To provide insights into this question, we turned to human expert explainers in the real-time strategy domain -- ``shoutcasters'' -- to understand (1) how they foraged in an evolving strategy game in real time, (2) how they assessed the players' behaviors, and (3) how they constructed pertinent and timely explanations out of their insights and delivered them to their audience.
The results provided insights into shoutcasters' foraging strategies for gleaning information necessary to assess and explain the players; a characterization of the types of implicit questions shoutcasters answered; and implications for creating explanations by using the patterns and abstraction levels these human experts revealed. 
\end{abstract}

\category{H.1.2}{User/Machine systems}{Human Information Processing} 
\category{H.5.2} {Information interfaces and presentation (e.g., HCI)} {User Interfaces}
\category{I.2.m}{Artificial Intelligence}{Miscellaneous}.

\keywords{\plainkeywords}

\input{1-Introduction}
\input{2-Background}

\input{3-Methodology}

\section{Results}
\input{4-ResultsPatches}
\input{5-ResultsNav}
\input{6-ResultsLimDey}

\input{7-ResultsContent}

\input{8-Conclusion}

\balance{}
\bibliographystyle{SIGCHI-Reference-Format}

\bibliography{references}

\end{document}

%% file: z-commands.tex
\newcommand{\boldify}[1]{}
\newcommand{\FIXME}[1]{}

\newcommand{\Ccode}[1]{{			\color{blue}
\textsl{#1}}}

\newcommand{\LimDey}[1]{{			
\texttt{#1}}}

\newcommand{\quotateInset}[2]{\vspace{-5pt}\begin{displayquote}						
Pair #1: \emph{#2}\end{displayquote}\vspace{-5pt}}

\newcommand{\quotate}[2]{{				
Pair #1: \emph{``#2.''}}}

\def\ImplicationsText{Implications for an interactive explainer}

\newcommand{\crazyPerformance}{
	\begin{sideways}
    \begin{minipage}{140pt}
    \centering 
    {\emph{Performance}}
	\end{minipage}\end{sideways}}
    
\newcommand{\crazyEnvironment}{
	\begin{sideways}\begin{minipage}{42pt}
    {\emph{\small Environment}}
	\end{minipage}\end{sideways}}
    
    \newcommand{\crazyActuator}{
	\begin{sideways}\begin{minipage}{30pt}
    {\emph{\small Actuator}}
	\end{minipage}\end{sideways}}
    
    \newcommand{\crazySensor}{
	\begin{sideways}\begin{minipage}{28pt}
    {\emph{\small Sensor}}
	\end{minipage}\end{sideways}}

\makeatletter
\def\url@leostyle{%
  \@ifundefined{selectfont}{
    \def\UrlFont{\sf}
  }{
    \def\UrlFont{\small\bf\ttfamily}
  }}
\makeatother
\def\pprw{8.5in}
\def\pprh{11in}

%% file: 1-Introduction.tex
\section{Introduction}

\boldify{Real-time strategy games are useful for AI research}

Real-time strategy (RTS) games are becoming more popular artificial intelligence (AI) research platforms.
A number of factors have contributed to this trend.
First, RTS games are a challenge for AI because they involve real-time adversarial planning within sequential, dynamic, and partially observable environments~\cite{ontanon}.
Second, AI advancements made in the RTS domain can be mapped to real world combat mission planning and execution such as an AI system trained to control a fleet of drones for missions in simulated environments~\cite{sycara2015abstraction}.

\boldify{But, the inner workings of the AI systems aren't easily understood by people such as our test pilot}

\input{figures/FigShoutcasters}

People without AI training will need to \emph{understand} and ultimately \emph{assess} the decisions of such a system, based on what such intelligent systems recommend or decide to do on their own.
For example, imagine ``Jake,'' the proud owner of a new self-driving car, who needs to monitor the AI system driving his car through complex traffic like the Los Angeles freeway system at rush hour, assessing when to trust the system~\cite{levineCarTrust} and when he should take control.
Ideally, an interactive explanation system could help Jake assess whether and when the AI is making its decisions ``\emph{for the right reasons}'' --- in real time.


\boldify{Hence Explainable AI, which we'll look at via StarCraft, which is very relevant to it}

Scenarios like this are the motivation for an emerging area of research referred to as ``Explainable AI,'' where an automated explanation device presents an AI system's decisions and actions in a form useful to the intended audience --- here, Jake.
There are recent research advances in explainable AI, as we discuss in the Related Work section, but only a few focus on explaining \emph{complex strategy environments} like RTS games and fewer draw from expert explainers.
To help fill this gap, we conducted an investigation in the setting of StarCraft II, a popular RTS game~\cite{ontanon} available to AI researchers~\cite{vinyals}.

\boldify{Shoutcasters are a good unit of analysis for two reasons. First, they face a black box TESTING task. Second, shoutcasters are EXPERT EXPLAINERS.}

We looked to ``shoutcasters'' (sportscasters for e-sports like RTS games) like those in Figure~\ref{fig:shoutcasters}.
In StarCraft e-sports, two players compete while the shoutcasters provide \emph{real-time} commentary.
They are helpful to investigate for explaining AI agents in real time to people like Jake for two reasons.
First, they face an \emph{assessment} task similar to explaining Jake's car-driving agent to him.
Specifically, they must 1) discover the actions of the player, 2) make sense of them and 3) assess them, particularly if they discover good, bad, or unorthodox behavior.
They must do all this while simultaneously constructing an explanation of their discoveries in real-time.

Second, shoutcasters are \emph{expert explainers}.
As communication professionals, they are paid to inform an audience they cannot see or receive feedback/questions from.
Hoffman \& Klein~\cite{hoffman2017explaining} researched five stages of explanation, looking at how explanations are formed from observation of an event, generating one or more possible explanations, judging the plausibility of said explanations, and either resolving or extending the explanation.
Their findings help to illustrate the complexity of shoutcasters' task, due to its abductive nature of explaining the past and anticipating the future.
In short, shoutcasters must anticipate and \emph{answer the questions the audience are not able to ask}, all while passively watching the video stream.

\boldify{Shoutcasters navigate an information interface to learn about the game state AND present it to the viewer, so we used IFT to understand it}

Because shoutcasters explain in parallel to gathering their information, we guided part of our investigation using Information Foraging Theory (IFT)~\cite{pirolli2007information}, which explains how people go about their information seeking activities.
It is based on naturalistic predator-prey models, in which the \emph{predator} (shoutcaster) searches \emph{patches} (parts of the information environment) to find \emph{prey} (evidence of players' decision process) by following the \emph{cues} (signposts in the environment that seem to point toward prey) based on their \emph{scent} (predator's guess at how related to the prey a cue is).
IFT constructs have been used to explain and predict people's information-seeking behavior in several domains, such as understanding navigations through web sites or programming and software engineering environments~\cite{chi2001using,fleming2013information,fu2007snif,kuttal2013predator,niu2013departures,perez2014diagnosis, piorkowski2015fix, piorkowski2016foraging,srinivasa2016foraging}.
However, to our knowledge, it has not been used before to investigate explaining RTS environments like StarCraft.

\boldify{Now you know the big picture, what are the RQs?}
Using this framework, we investigated the following research questions (RQs):
\vspace{-5pt}
\begin{enumerate}[labelindent=20pt,labelwidth=\widthof{\ref{last-item}},label=\arabic*.,itemindent=1em,leftmargin=!]
\item[\textbf{RQ1}] \emph{The What and the Where}: What information do shoutcasters seek to generate explanations, and where do they find it?
\vspace{-5pt}
\item[\textbf{RQ2}] \emph{The How}: How do shoutcasters seek the information they seek?
\vspace{-5pt}
\item[\textbf{RQ3}] \emph{The Questions}: What implicit questions do shoutcasters answer and how do they form their answers?
\vspace{-5pt}
\item[\textbf{RQ4}] \emph{The Explanations}: What relationships and objects do shoutcasters use when building their explanations? \label{last-item}
\end{enumerate}
\vspace{-5pt}

%% file: figures/FigShoutcasters.tex
\begin{figure}[b!]
	\centering
	\includegraphics[width=.67\linewidth]{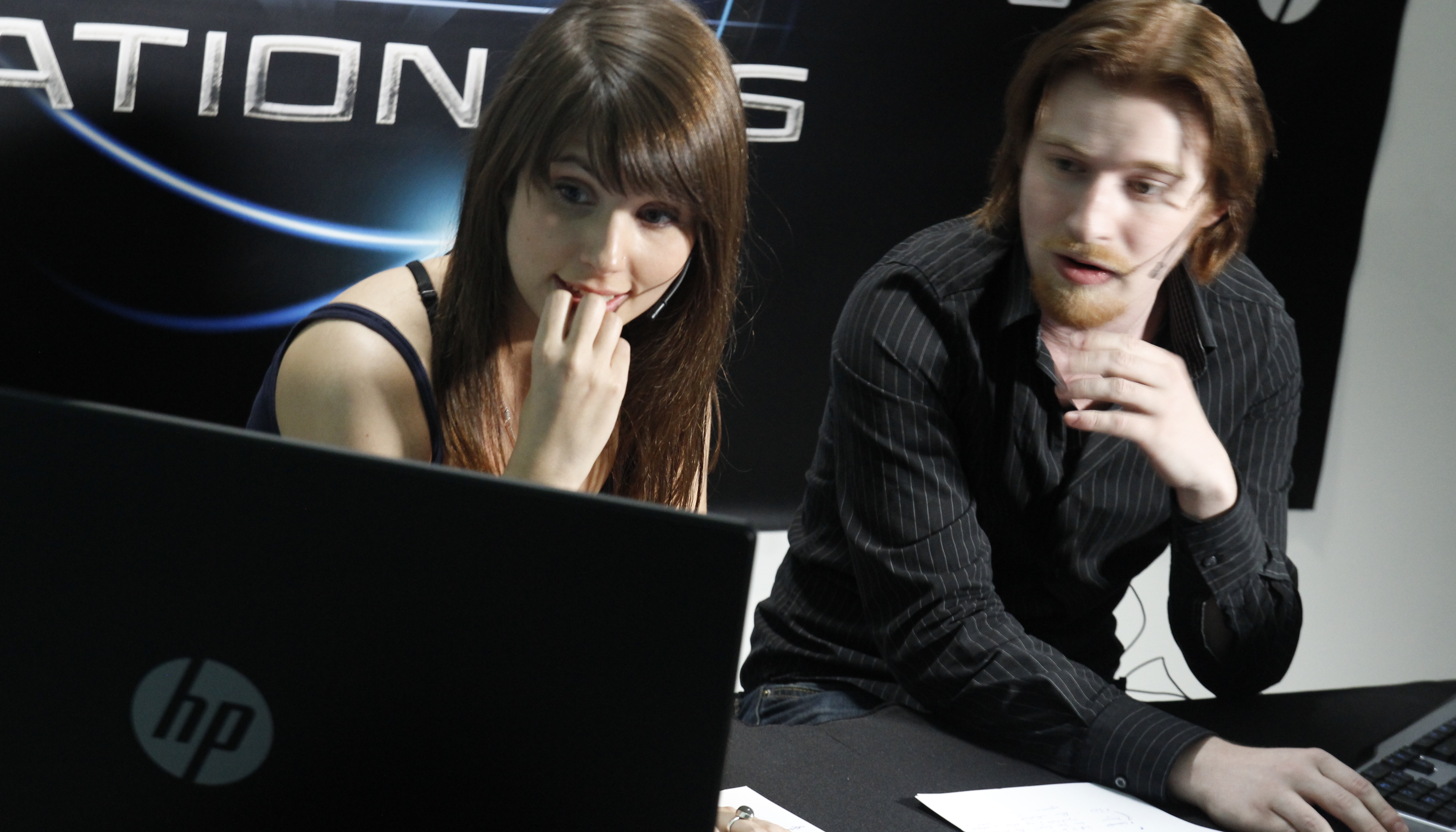}
	\caption{Two shoutcasters providing commentary for a professional StarCraft match.}
	\label{fig:shoutcasters}
\end{figure}

%% file: 2-Background.tex
\section{Background and Related Work}

\boldify{mental models: Kulesza found that Action/prediction explanation builds trust in the system}

In the HCI community, research has begun to investigate the benefits to humans of explaining AI.
``Jake'' (our test pilot) improving his mental model is critical to his success, since Jake needs a reasonable mental model of the AI system to assess whether it is making decisions \emph{for the right reasons}.
Mental models, defined as ``internal representations that people build based on their experiences in the real world,'' enable users to predict system behavior~\cite{norman1983some}.
Kulesza et al.~\cite{kulesza2012tell} found those who adjusted their mental models most in response to explanations of AI (a recommender system) were best able to customize recommendations.
Further, participants who improved their mental models the most found debugging more worthwhile and engaging.

Building upon this finding, Kulesza et al.~\cite{kulesza2015principles} then identified principles for explaining (in a ``white box'' fashion) to users how a machine learning system makes its predictions more transparent to the user.
In user studies with a prototype following these principles, participants' quality of mental models increased by up to 52\%, and along with these improvements came better ability to customize the intelligent agents.
Kapoor et al.~\cite{kapoor2010interactive} also showed that explaining AI increased user satisfaction and interacting with the explanations enabled users to construct classifiers that were more aligned with target preferences.
Bostandjiev et al.'s work on a music recommendation system~\cite{bostandjiev2012tasteweights} found that explanation led to a remarkable increase in user-satisfaction with their system.

\boldify{One of the pioneers in explainable AI, Lim identified that are important to answer in order to explain an intelligent system.}

As to what people \emph{want} explained about AI systems, one influential work into explainable AI has been Lim \& Dey's~\cite{lim2009} investigation into information demanded from context-aware intelligent systems.
They categorized users' information needs into various ``intelligibility types,'' and investigated which types provided the most benefit to user understanding.
Among these types were ``What'' questions (What did the system do?), ``Why'' questions (Why did the system do X?), and so on.
In this paper, we draw upon these results to categorize the kinds of questions that shoutcasters' explanations answered.

Other research confirms that explanations containing \emph{certain} intelligibility types make a difference in user attitude towards the system.  
For example, findings by Cotter et al.~\cite{cotter2017explaining} showed that justifying
\emph{why} an algorithm works (but not on \emph{how} it works) were helpful for increasing users' confidence in the system --- but not for improving their trust.
Other work shows that the relative importance of the intellibility types may vary with the domain; for example, findings by Castelli et al.~\cite{castelli2017happened} in the domain of smart homes showed a strong interest in ``What'' questions, but few of the other intellibility types.

\input{tables/TableMatchSummary}

\boldify{However, it's hard, and sometimes there seem to be tradeoffs.}

Constructing effective explanations of AI is not straightforward, especially when the underlying AI system is complex.  
Both Kulesza et al.~\cite{kulesza2015principles} and Guestrin et al.~\cite{ribeiro2016should} point to a potential trade-off between \emph{faithfulness} and \emph{interpretability} in explanation. 
The latter group developed an algorithm that can explain (in a ``black box'' fashion) predictions of any classifier in a faithful way, and also approximate it locally with an interpretable model.
In their work, they described the fidelity-interpretability trade-off, in which making an explanation more faithful was likely to reduce its interpretability, and vice versa.
However, humans manage this trade-off by accounting for many factors, such as the audience's current situation, their background, amount of time available, etc.
One goal of the current study is to understand how expert human explainers, like our shoutcasters, manage this trade-off.

\boldify{In the domain of RTS...}
In the domain of assessing RTS intelligent agents, Kim et al.~\cite{kim2016evaluation} invited 20 experienced players to assess the skill levels of AI bots playing StarCraft. 
They observed that human rankings were different in several ways to a ranking computed from the bots' competition win rate, because humans weighed certain factors like decision-making skill more heavily.
The mismatch between empirical results and perception scores may be because AI bots that are effective against each other proved less effective against humans.


Cheung et al.~\cite{Cheung:2011:SSU:1978942.1979053} studied StarCraft from a different perspective, that of non-participant spectators.
Their investigations produced a set of nine personas that helped to illuminate \emph{who} these spectators are and \emph{why} they watch.
Since shoutcasters are one of the personas, they discussed how shoutcasters affect the spectator experience and how they judiciously decide how and when to reveal different types of information, both to entertain and inform the audience.

\boldify{The closest to our own work is Metoyer's in the RTS domain}

The closest work to our own is Metoyer et al.'s~\cite{metoyer2010explaining} investigation into the vocabulary and language structure of explaining RTS games.
In their study, novices and experts acted in pairs, with the novice watching the expert play and providing questions, while the expert thought aloud and answered questions.
They developed qualitative coding schemes of the content and structure of the explanations the expert players offered.
Our investigation is subtly different in that our explainers are expert \emph{communicators} about the game and must \emph{anticipate} audience questions on their own.
However, given the pertinence of their work, we modified their code set to analyze shoutcasters' utterances.

%% file: tables/TableMatchSummary.tex
\begin{table*}
	\centering
	\begin{tabular}{l| llll}
	   &\textbf{Tournament} & \textbf{Shoutcasters} & \textbf{Players} & \textbf{Game}\\
	   \hline
	1 & 2017 IEM Katowice & ToD and PiG & Neeb vs Jjakji & 2\\
	2 & 2017 IEM Katowice & Rotterdam and Maynarde & Harstem vs TY & 1\\
	3 & 2017 GSL Season 1 Code S & Artosis and tasteless & Soo vs Dark & 2\\
	4 & 2016 WESG Finals & Tenshi and Zeweig & DeMuslim vs iGXY & 1\\
	5 & 2017 StarLeague S1 Premier& Wolf and Brendan & Innovation vs Dark & 1 \\
	\hline
	6 &2016 KeSPA Cup & Wolf and Brendan & Maru vs Patience & 1\\
	7 & 2016 IEM Geonggi & Kaelaris and Funka & Byun vs Iasonu & 2\\
	8 & 2016 IEM Shanghai & Rotterdam and Nathanias & ShowTime vs Iasonu & 3\\
	9 & 2016 WCS Global Finals & iNcontroL and Rotterdam & Nerchio vs Elazer & 2\\
	10 & 2016 DreamHack Open Leipzig & Rifkin and ZombieGrub & Snute vs ShowTime & 3\\
	\end{tabular}
	\caption{Summary of StarCraft 2 games studied. Please consult our supplementary materials for transcripts and links to videos.
    }
	\label{table:matchSummary} 
\end{table*}

%% file: 3-Methodology.tex
\section{Methodology}

\input{figures/FigGameScreenshot}

\boldify{The corpus consists of videos casted by top tier shoutcasters. This was done because we wanted to ensure high quality explanation and agent performance.}

In order to study high quality explanations and capable players, we considered only games from professional tournaments denoted as ``Premier'' by TeamLiquid\footnote{\url{http://wiki.teamliquid.net/starcraft2/Premier_Tournaments}}.
Using these criteria, we selected 10 matches available with video on demand from professional StarCraft 2 tournaments from 2016 and 2017 (Table~\ref{table:matchSummary}).
Professional matches have multiple games, 
so we randomly selected one game from each match for analysis.
16 distinct shoutcasters appeared across the 10 videos, with two casters\footnote{Here, caster pair (\emph{caster} or \emph{pair} for short) differentiates our observed  individuals from the population of shoutcasters as a whole.} commentating each time.


\boldify{We first did the Explanation coding on the corpus, with acceptable IRR achieved. Why: because the cast is done for entertainment value as well as information, so it needed filtration.}

Shoutcasters should both inform and \emph{entertain}, so they fill dead air time with jokes.
Thus, we filtered the casters' utterances by relevance.
To do so, two researchers independently coded 32\% of statements in the corpus as relevant or irrelevant to explaining the game.  
We achieved a 95\% inter-rater reliability (IRR), as measured by the Jaccard index.
(The Jaccard index is the size of the intersection of the codes applied by the researchers divided by the size of the union.) 
Then, the researchers split up and coded the rest of the corpus. 

\boldify{RQ1 and RQ2 about about IFT, so that guided our analysis of these two RQs.  For RQ1 we didn't have to code. For RQ2, we coded...} 

Research questions RQ1 and RQ2 investigated how the casters seek information onscreen, so we used IFT constructs to discover the types of information casters sought and how they unearthed it.
For RQ1 (the patches in which they sought information), we simply counted the casters' navigations among patches.  
Changes in the display screen identified these for us automatically.
For RQ2 (\emph{how} they went about their information foraging), we coded the 110 instances of caster navigation by the context where it took place, based on player actions --- Building, Fighting, Moving, Scouting --- or simply caster navigation.
Two researchers independently coded 21\% of the data in this manner, with IRR of 80\% (Jaccard). 
After achieving IRR, one researcher coded the remainder of the data.

\boldify{for RQ3, Lim/Dey coding was performed on our modified coding rules, with sufficient IRR achieved. This was done to try to understand HOW the explanation proceeds}

For RQ3 (implicit questions the shoutcasters answered), we coded the casters' utterances by the Lim \& Dey~\cite{lim2009} questions they answered.
We added a judgment code to capture caster evaluation on the \emph{quality} of actions. 
The complete code set will be detailed in the RQ3 Results section.
Using this code set, two researchers independently coded 34\% of the 1024 explanations in the corpus, with 80\% inter-rater reliability (Jaccard).
After achieving IRR, the researchers split up the remainder of the coding. 

\boldify{for RQ4, Content coding was performed on the coding rules taken from Metoyer et al. We did this because we wanted to know WHAT the explanations contained. Coding rules were modified slightly to better fit the domain of StarCraft}

To investigate RQ4 (explanation content), 
we drew content coding rules from Metoyer et al.~\cite{metoyer2010explaining}'s analysis of explaining Wargus games and added some codes to account for differences in gameplay and study structure.
(For ease of presentation, in this paper we use the terms ``numeric quantity'' and ``indefinite quantity'' instead of their terms ``identified discrete'' and ``indefinite quantity'', respectively.)
Two researchers independently coded the corpus, one category at a time (e.g., Objects, Actions, ...), achieving an average of 78\% IRR on more than 20\% of the data in each category.
One researcher then finished coding the rest of the corpus. 
Since all data sources are public, we have provided all data and coding rules in supplementary materials to enable replicability and support further research.

%% file: figures/FigGameScreenshot.tex
\begin{figure*}
	\centering
	\includegraphics[width=.9\textwidth]{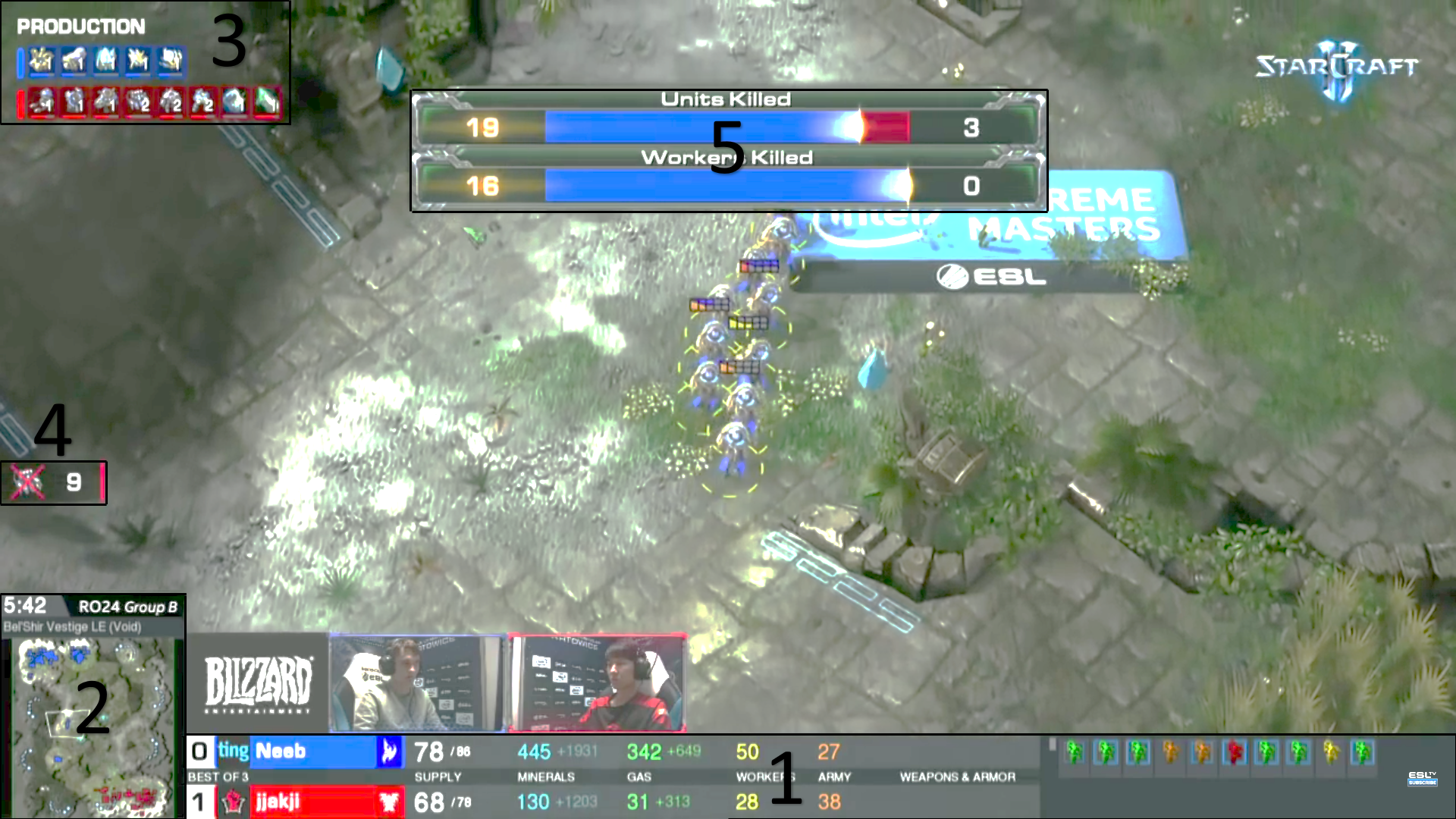}
	\caption{
    A screenshot from an analyzed game, modified to highlight the patches available to our casters: 
\emph{HUD} [1, bottom] (Information about current game state, e.g., resources held, income rate, supply, and upgrade status); 
\emph{Minimap} [2, lower left] (Zoomed out version of the main window);
\emph{``Tab''} [3, top left] (Provides details on demand,  currently set on ``Production'');
\emph{Workers killed} [4, center left] (Shows that 9 Red workers have died recently);
\emph{Popup} [5, center] (visualizations that compare player performance, usually shown briefly).
Regions 3 and 5 will be detailed in Figures~\ref{fig:performance} and \ref{fig:productionTab}.
}
	\label{fig:gameScreenshot}
\end{figure*}

%% file: 4-ResultsPatches.tex
\subsection{RQ1 Results: What information do shoutcasters seek to generate explanations, and where do they find it?}

\boldify{We will use two frameworks to understand shoutcasters info seeking: PEAS to understand the WHAT, and IFT to understand the WHERE and HOW of  shoutcasters' info seeking.}

We used two frameworks to investigate casters' information seeking behaviors.
To situate \emph{what} information casters sought in a common framework for conceptualizing intelligent agents, we turned to the Performance, Environment, Actuators, Sensors (PEAS) model~\cite{Russell:2003:AIM:773294}. 
We drew from Information Foraging Theory (IFT) to understand \emph{where} casters did their information seeking, beginning with the places their desired information could be found. 
These places are called information ``patches'' in IFT terminology.  

\boldify{Performance measures are...<defn>, and are interesting because they aren't used much}

Table~\ref{table:IFTpatchNav} columns 1 and 2 show the correspondence between PEAS constructs and patches in the game that the casters in our data actually used.
\textbf{Performance} measures showed assets, resources, successes, and failures, e.g., Figure~\ref{fig:gameScreenshot} region 4 (showing that Blue has killed 9 of Red's workers) and region 5 (showing that Blue has killed 19 units to Red's 3, etc.).
Table~\ref{table:IFTpatchNav} shows that casters rarely consulted performance measures, especially those that examined \emph{past} game states.
However, they discussed basic performance measures available in the HUD (Figure~\ref{fig:gameScreenshot} region 1), which contained \emph{present} state information, e.g., resources held or upgrade status.

\input{tables/TableIFTpatchNav}

\boldify{Environment is... and is essentially not interesting for us here}

The \textbf{Environment}, or where the agent is situated, corresponds to the game state (map, units, structures, etc.), shown in the main window in Figure~\ref{fig:gameScreenshot}.
The Environment mattered for ambient awareness, 
but casters navigated to Sensor and Actuator information most actively, so we turn to those constructs next.

\boldify{Sensors are..., and are interesting because...}

\textbf{Sensors} helped the agent collect information about the environment and corresponded to the local vision area provided by individual units themselves in our domain.
Figure~\ref{fig:gameScreenshot} region 2 (Minimap) shows a ``bird's eye view'' of the portion of the environment observable by the Sensors.
Casters used patches containing information about Sensors very often, with Minimap and Vision Toggles being among the most used patches in Table~\ref{table:IFTpatchNav}. 
The casters had ``superpowers'' with respect to Sensors (and performance measures) --- their interface allowed \emph{full} observation of the environment, whereas players could only \emph{partially} observe it.
As the only ways for casters to peer through the players' sensors, the casters extensively used the Minimap and the Vision Toggle.

\boldify{Actuators are..., and are interesting because...}

\textbf{Actuators} were the means for the agents to interact with their environment, such as building a unit.
Figure~\ref{fig:gameScreenshot} region 3 (Production Tab) shows some of the actuators the player was using, namely that Player Blue was building 5 types of objects, whereas Red was building 8.
Casters almost always kept visualizations of actions \emph{in progress} on display.
RTS actions had a \emph{duration}, meaning that when a player took an action, time passed before its consequence had been realized.
The Production tab's popularity was likely due to the fact that it is the \emph{only stable} view of information about actuators and their associated actions.

\boldify{In their own words, actuators are interesting! E.g.}

In fact, prior to the game in our corpus, Pair 3 had this exchange, which demonstrated their perception of the production tab's importance to doing their job, 
\quotateInset{3a}{``What if we took someone who knows literally nothing about StarCraft, just teach them a few phrases and \emph{\textbf{what everything is on the production tab}}?'' \\\emph{Pair 3b:}``Oh, I would be out of a job.''}

\subsubsection{\ImplicationsText}

\boldify{"keep your sensors close, but your actuators closer,"! }

Abstracting beyond the StarCraft components to the PEAS model revealed a pattern of the casters' behaviors with implications for future explanation systems, which we characterize as:
``keep your Sensors close, but your Actuators closer.''
This aligns with other research, showing that real-time visualization of agent actions can improve system transparency~\cite{WorthamRobots}.

However, these results contrast with many of today's explanation systems, which tend to prioritize Performance measures, but the Actuators and Sensors seemed to form the very core of these expert explainers' information seeking for presentation to their audience.
Our results instead suggest that an explanation system should prioritize useful, readily accessible information about what an agent did or can do (Actuators) and of what it can see or has seen (Sensors).

%% file: tables/TableIFTpatchNav.tex
\begin{table*}
	\centering
	\begin{tabular}{@{}l|p{.35\linewidth}| ll|l|lllll lllll@{}}
    & \textbf{Patch Name} & \textbf{State} & \textbf{Agg.} &   
    \begin{sideways}Usage\end{sideways} & 
    \begin{sideways}Pair 1\end{sideways} &
    \begin{sideways}Pair 2\end{sideways} &
    \begin{sideways}Pair 3\end{sideways} &
    \begin{sideways}Pair 4\end{sideways} &
    \begin{sideways}Pair 5\end{sideways} &
    \begin{sideways}Pair 6\end{sideways} &
    \begin{sideways}Pair 7\end{sideways} &
    \begin{sideways}Pair 8\end{sideways} &
    \begin{sideways}Pair 9\end{sideways} &
    \begin{sideways}Pair 10\end{sideways}  \\
	   \hline
       
 \multirow{7}{*}{\crazyPerformance}
 
         &\textbf{Units Lost popup}: Shows count and resource value of the units each player has lost.
        & Past		& High	 & 6 
        &  &  &  &  &  & 2 & 2 & 1 & 1 &   \\ \cline{2-15}
        
        &\textbf{Units Lost tab}: Same as above, but as a tab.
        & Past		& High		& 5
        & 1 & 1 &  &  & 1 & 1 & 1 &  &  &   \\ \cline{2-15}        
        
        &\textbf{Income tab}: Provides resource gain rate.
        & Present & High  & 2
        &  &  &  & 1 &  & 1 &  &  &  &   \\ \cline{2-15} 
        
        &\textbf{Income popup}: Shows each player's resource gain rate and worker count.
        & Present & High  & 2
    	&  &  &  &  & 1 & 1 &  &  &  &   \\ \cline{2-15}          
        
        &\textbf{Army tab}: Shows supply and resource value of currently held non-worker units.
        & Present & High  & 1
        &  &  &  & 1 &  &  &  &  &  &  \\ \cline{2-15} 
        
        &\textbf{Income Advantage graph}: Shows time series data comparing resource gain rate.
        & Past & High  & 1
        & 1 &  &  &  &  &  &  &  &  &   \\ \cline{2-15} 
        
        &\textbf{Units Killed popup}: Essentially the opposite of the Units Lost popup
        & Past & High  & 1
        & 1 &  &  &  &  &  &  &  &  &   \\ \hline

 \multirow{3}{*}{\crazyEnvironment} 
    
            &\textbf{Units tab:} Shows currently possessed units.		
        & Present	& Low	 & 51
    	& 1 & 2 & 2 & 10 & 1 & 13 & 20 & 2 &  &   \\ \cline{2-15}  
    
         &\textbf{Upgrades tab}: Like Units tab, but for upgrades to units and buildings.
       & Present	& Low	 & 5
        &  &  &  & 1 & 3 &  &  & 1 &  &   \\ \cline{2-15} 
        
        &\textbf{Structures tab}: Like Units tab, but buildings.
        & Present	& Low	 & 2
        & 1 &  &  & 1 &  &  &  &  &  &   \\ \hline

\multirow{1}{*}{\crazyActuator}    
    
       &\textbf{Production tab}:  Shows the units, structures, and upgrades that are in progress, i.e. have been started, but are yet to finish.
       & Present & Low 
       & \multicolumn{11}{c}{Preferred (by choice) ``always-on'' tab, not counted} \\ \hline

  \multirow{2}{*}{\crazySensor}          
       
       &\textbf{Minimap:} Shows zoomed out map view
        & Present & Med 
       & \multicolumn{11}{c}{Too many to count} \\ \cline{2-15} 
     	
        &\textbf{Vision Toggle:} Shows only the vision available to \emph{one} of the players.
        & Present 	& Low		& 36
        & 5 & 8 & 1 & 2 & 1 & 5 & 1 & 7 & 5 &  1  \\
	\end{tabular}
    
	\caption{This table illustrates description, classification, and usage rates of the patches and enrichment operations we observed casters using.  
    Each patch is classified by:
    1. The part of the PEAS model that this patch illuminates best (column 1),
    2. whether it examines past or present game states (column 3), and
    3. degree to which the patch aggregates data in its visualization (column 4).
    The remaining columns show total usage counts, as well as per caster pair usage.
    Note that there are additional patches passively available (Main Window and HUD) which do not have navigation properties.
    }
	\label{table:IFTpatchNav} 
\end{table*}

%% file: 5-ResultsNav.tex
\subsection{RQ2 Results: The How: How do shoutcasters seek the information they seek?}

\boldify{AEP+S appears to be the order the casters care about patches AND navigate}

Section RQ1 discussed the What and Where (i.e., the content casters sought and locations where they sought it.)
We then considered how they decided to move among these places.   

The casters' foraging moves seemed to follow
a common foraging ``loop'' through the available information patch types: an Actuators-Environment-Performance loop with occasional forays over to Sensors (Figure~\ref{fig:strategyIFT}).
Specifically, the casters tended to start at  
the ``always-on'' Actuator-related patches of current state's \emph{actions in-progress};
then when something triggered a change in their focus, they checked the Environment for current game state information  
and occasionally Performance measures of past states. 
If they needed more information along the way, they went to the Sensors to see through a player's eyes.
We will refer to this process as the ``A-E-P+S loop''.

\input{figures/FigStrategyIFT}

\boldify{Why move? Well, IFT explains how people make decisions to move from one place to another.}

To help derive what caused casters to leave Actuator patches, which seemed to have so much importance to them,  
Information Foraging Theory (IFT) explains why people (information predators) leave one patch to move to another as a cost/benefit decision, based on the value of information in the patch a predator is already in versus the value per cost of going to another patch~\cite{pirolli2007information}.  
Staying in the same patch is generally the least expensive, but when there is less value to be gained by staying versus moving to another patch, the predator moves to the other patch.
However, the predator is not omniscient: decisions are based upon the predator's \emph{perception} of the cost and value that other patches will actually deliver. 
They formed these perceptions from both their prior experience with different patch types~\cite{piorkowski2015fix} and from the cues (signposts in their information environment) that provided concise information about content available in other patches.
For the casters, certain types of cues tended to trigger a move.

\input{figures/FigPerformance}

\boldify{(PEAS "E") Cue 1: when COMBAT occurred they went to ENVIRONMENT }

Impending combat was the most common cue triggering a move from the Actuators type (Production tab) to the Environment type (Units tab) --- i.e., from A to E in the A-E-P+S loop.
In Figure~\ref{fig:strategyIFT}, the cue was co-located, opposing units, indicative of imminent combat, which led to caster navigation to a new patch to illuminate the environment.
In fact, combat cues triggered  navigations to the Units tab most frequently, accounting for 30 of the 51 navigations there
(Table~\ref{table:IFTpatchNav}).

Interestingly, this cue type was different from the static cues most prior IFT research has used.  
In previous IFT investigations, cues tended to be static decorations (text or occasionally images) that label a navigation device, like a hyperlink or button that leads to another information patch.  
In contrast, cues like the onset of combat are dynamic and often did not provide an affordable direct navigation. 
However, cues like this were considered cues because they 
``provide users with concise information about content that is not immediately available''~\cite{pirolli2007information}.
In the case of combat, they suggested high value in another location, namely the Units tab.

%

\boldify{Cue 2: The end of combat leads to "P": some kind of postmortem analysis after some set of actions had completed}

Combat ending was a dynamic cue that triggered a move to a  Performance measure.
Of the 13 navigations to a past-facing Performance measure (via tab or popup), 10 occurred shortly after combat ended as a form of ``after-action review.''
Occasionally, the shoutcasters visited other Performance  patches, such as the Income, Units Lost, and Army tabs, to demonstrate reasons why a player had accrued an in-game lead, or the magnitude of that lead (7 navigations).
However, signs of completed fighting were the main cues for visiting a Performance patch.

\input{figures/FigProductionTab}

\boldify{(to S) (unknown) Cue 3: When an unknown cue happened, casters decided to assess scouting actions, leading to "S".  The casters have ALL vision, and use vision toggle to ``lose their superpowers'' and use the mortal's view to assess scouting actions}

The most common detour out of the A-E-P part of the loop to a Sensor patch was to enrich the information environment via the Vision Toggle (36 navigations).
The data did not reveal exactly what cue(s) led to this move, but the move itself had a common theme: to assess scouting operations.
The casters used the Vision Toggle to allow themselves to see the game through the eyes of only \emph{one} of the players, but their default behavior was to view the game with ALL vision. This provided the casters with the ability to observe \emph{both} players' units and structures simultaneously.
Toggling the Vision Sensor in this way enabled them to 
assess what information was or had been gathered by each player via their scouting actions (29 of the 36 total Vision Toggles), since an enemy unit would only appear to the player's sensors if they had a friendly unit (e.g., a scout) nearby.
Toggling the vision Sensor was the second most common patch move.

\boldify{(PEAS "P") They also enriched their environment by adding on information visualizations derived from the raw data, like Performance Measure popups or other basic visualizations.}

Besides the act of following cues, IFT has another foraging operation: \emph{enriching} their information environment to make it more valuable or cost-efficient~\cite{pirolli2007information}.
The aforementioned Vision Toggle was one example of this, and another was when casters added on information visualizations derived from the raw data, like Performance measure popups or other basic visualizations.
Two examples of the data obtained through this enrichment is shown  in Figure~\ref{fig:performance}.


\boldify{Performance measures mattered because of their data aggregation}

These Performance measures gave the shoutcasters at-a-glance information about the ways one player was winning.
For example, the most commonly used tab, Units Lost tab (Figure~\ref{fig:performance}) showed the number of units lost and their total value, in terms of resources spent.
This measure achieves ``at a glance'' by aggregating \emph{all} the data samples together by taking a \emph{sum}; derived values like this allow the visualization to scale to large data sets~\cite{Spence:2007:IVD:1196684}.
However, Table~\ref{table:IFTpatchNav} indicates that the lower data aggregation patches were more heavily used.
As Figure~\ref{fig:productionTab} shows, the casters used the Production tab to see units grouped by type, so \emph{type} information was maintained with only \emph{positional} data lost.
This contrasts with the Minimap (medium aggregation), in which type information is discarded but positional information maintained at a lower \emph{granularity}.
The casters used Performance measure patches primarily to understand present state data (HUD), but these patches were also the only way to access \emph{past} state information (Table~\ref{table:IFTpatchNav}).

\subsubsection{\ImplicationsText}

\boldify{moving thru PATCHES: AESP appears to be the order the casters care AND navigate) 
}

These results have several implications for automated explanation systems in this domain.
First, the A-E-P+S loop and how the casters traversed it reveals priority and timing implications for automated explanation systems. 
For example, the cues that led them to switch to different information patches could also be cues in an automated system about the need to avail different information at appropriate times.
For example, our casters showed a strong preference for actuator information as ``steady state'' visualization, but preferred performance information upon conclusion of a subtask.

\boldify{PEAS+IFT: Regarding (CUES) and (PATCHES), currently they can only get at fighting or scouting actions via careful placement of the camera. Minimap navigations were basically uncountable, and the various tabs were also used to move the camera, but the system still breaks down...}

Viewing the casters' behaviors through the dual lens of PEAS + IFT has implications for not only the kinds of patches that an explanation system would need to provide, but also the cost to users of not providing these patches in a readily accessible format.
For example, PEAS + IFT revealed a costly foraging problem for the casters due to the relative inaccessibility of some Actuator patches.
In StarCraft, there is no easily accessible mechanism by which they could navigate to an Actuator patch with fighting or scouting actions in progress. 

Instead, the only way the casters could get access to these actions was via \emph{painstaking} camera placement.
To accomplish this, the casters made countless navigations to move the camera using the Minimap, traditional scrolling, or via tabs with links to the right unit or building.
But despite all these navigation affordances, sometimes the casters were unable to place the camera on all the actions they needed to see.  

For example, at one point when Pair 4 had the camera on a fight at Xy's base, a second fight broke out at DeMuslim's base, which they completely missed:

\quotateInset{4a}{<surprised, noticing something amiss> \\``Xy actually killed the 3rd base of DeMuslim.''
\\...<the pair tries to figure what must have happened>...
\\\emph{Pair 4b:} ``Oh my god, you're right Alex.'' 
\\\emph{Pair 4a:} ``Yeah, it was killed during all that action.''}

%% file: figures/FigStrategyIFT.tex
\begin{figure}
	\centering
    \includegraphics[width=\columnwidth]{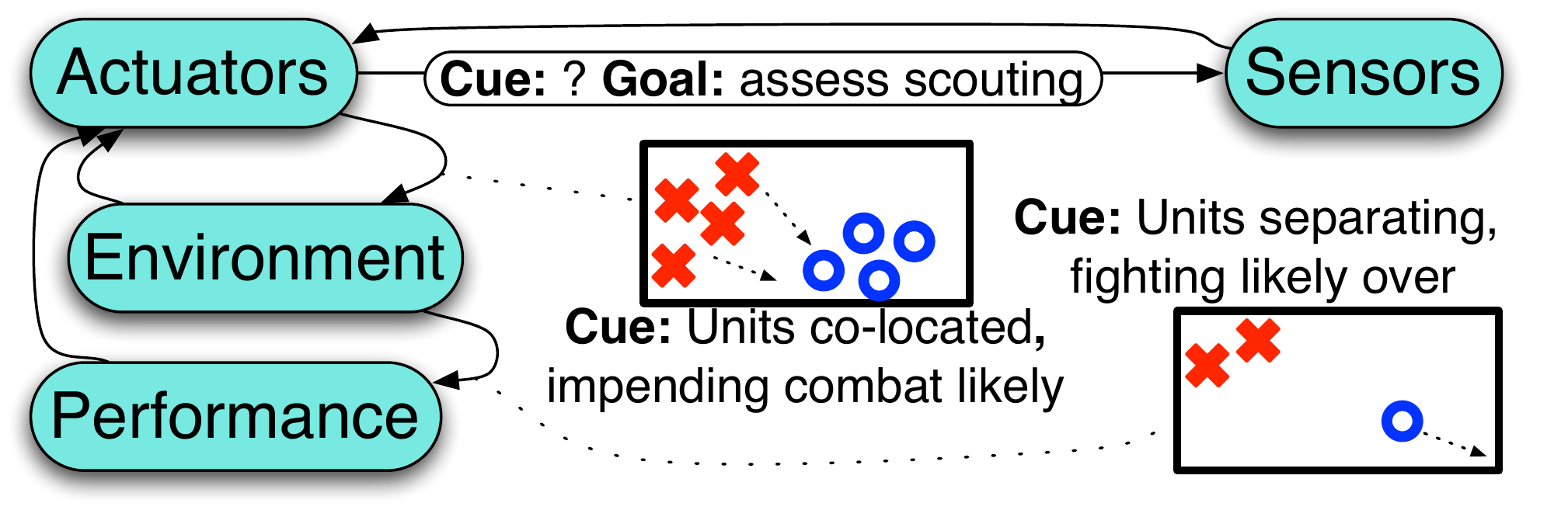}
    \caption{The A-E-P+S loop was a common information foraging strategy some casters used in foraging for agent behavior. 
    It starts at the Actuators, and returns there throughout the foraging process. 
    If a caster interrupted the loop, they usually did so to return to the Actuators.
    }
    \label{fig:strategyIFT}
\end{figure}

%% file: figures/FigPerformance.tex
\begin{figure}
	\centering
	\includegraphics[width=.37\linewidth]{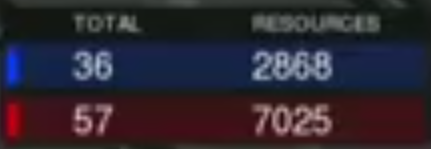}
    \hfill
\includegraphics[width=.60\linewidth]{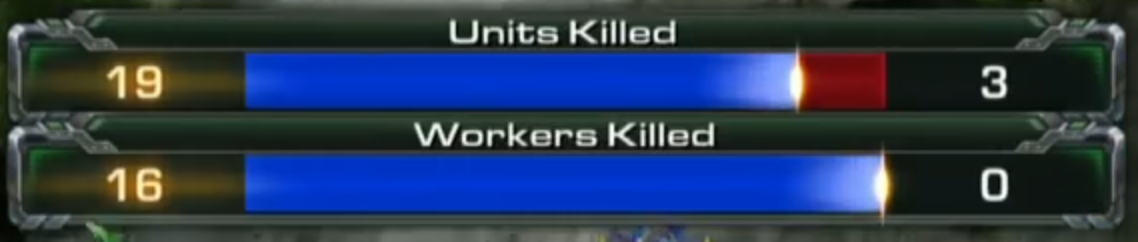}
	\caption{
    The Units Lost tab (left image) shows the number of units lost and their total value, in terms of resources spent, for both players.
    In this example from Pair 2, we see that Blue Player (top) has lost merely 2800 minerals worth of units so far in this game, while Red has lost more than 7000.
    The Units Killed popup (right image) allows shoutcasters to quickly compare player performance via a ``tug-of-war'' visualization.
    In this example from Pair 1, as we see that Blue Player (left) has killed 19 units, while Red has killed merely 3.
    The main difference between these two styles of visualization is that the tab offers more options and information depth to ``drill down'' into.
    }
	\label{fig:performance}
\end{figure}

%% file: figures/FigProductionTab.tex
\begin{figure}
	\centering
	\includegraphics[width=\linewidth]{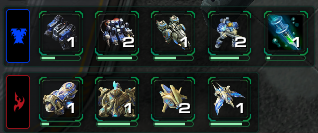}
	\caption{The Production tab, showing the build actions currently in progress for each player.
    Each unit/structure type is represented by a glyph (which serves as a link to navigate to that object), provided a progress bar for duration, and given the number of objects of that type.
    Thus, we can see that Blue Player (top row) is building 5 different types of things, while Red (bottom row) is building 4 types of things.
    The Structures, Upgrades, and Units tab look fairly similar to the Production tab.
    }
	\label{fig:productionTab}
\end{figure}

%% file: 6-ResultsLimDey.tex
\subsection{RQ3 Results: What implicit questions do shoutcasters answer and how do they form their answers?}

\input{figures/FigLimDeyFreq}

\input{tables/TableLimDeyCoding}


\boldify{The shoutcasters' foraging shows how THEY discovered information to make sense of things. Now, let's look at how they support sensemaking by others through their explanations (This is going to be PEAS.A.explain)}

For the first two research questions, we considered how the shoutcasters gathered and assessed information.
We now shift focus to the explanations themselves. 

Much of the prior research into explaining agent behavior starts at some kind of observable effect and then explains something about that effect or its causes~\cite{kapoor2010interactive, kulesza2015principles, lim2009assessing, tullio2007}.
In RTS games, most such observable effects are the result of player actions, and recall from RQ1 that the casters spent most of their information-gathering effort examining the players' Actuators to discover and understand actions. 

The casters used the information they gained to craft explanations to answer implicit questions (i.e., questions their audience ``should be'' wondering) about player actions. 
Thus, drawing from prior work about the nature of questions people ask about AI, we coded the 1024 casters' explanations using the Lim \& Dey ``intelligibility types''~\cite{lim2009}.

\boldify{Results from Lim coding were remarkably consistent across shoutcasters, and showed that the order (by frequency) of questions answered by shoutcasters was "What," "What could happen," "How to," "How good/bad was that action," "Why," and "Why didn't."}

The shoutcasters were remarkably consistent (Figure~\ref{fig:limDeyFreq}) in the types of implicit questions they answered. 
As Table~\ref{table:limDeyCoding} sums up, casters overwhelmingly chose to answer \LimDey{What}, with \LimDey{What-could-happen} and \LimDey{How-to} high on their list.
(The total is greater than 1024 because explanations answered multiple questions and/or fit into multiple categories.)

\boldify{This is a surprise! Very different from Lim \& Dey results. They observe high user demand of why, we observe very little delivery of it by explainers.}

These results surprised us.
Whereas Lim \& Dey~\cite{lim2009assessing} found that \LimDey{Why} was the most demanded explanation type from users, the casters rarely provided \LimDey{Why} answers.
More specifically, in the Lim \& Dey study, approximately 48 of 250 participants, (19\%) demanded a \LimDey{Why} explanation.
To contrast with our study, only 27 of the casters' 1024 utterances (approximately 3\%) were \LimDey{Why} answers.

\subsubsection{Discussion and implications for an interactive explainer}

Why so few \LimDey{Why}s? 
Should an automated explainer, like our shoutcasters, eschew \LimDey{Why} explanations, in favor of \LimDey{What}?

\boldify{Audience of shoutcasted videos are likely well informed, thus expectations are met by receiving just the play-by-play. }

One possibility is that the casters delivered exactly what their audience wanted, and thus the casters' distribution of explanation types was well chosen.  
After all, the casters were experts paid to provide commentary for prestigious tournaments, so they would know their audience well. 
The expertise level of the audience may have been fairly high, because the tournament videos were available only \emph{on demand} (as opposed to \emph{broadcast} like some professional sports) at websites that casual audience members may not even know about.
If a well-informed audience expected the players to do exactly what they did, 
their expectations would not be violated, which, according to Lim \& Dey, suggests less demand for \LimDey{Why}~\cite{lim2009assessing}.
This suggests that the extent to which an automated explainer needs to emphasize 
\LimDey{Why} explanations may depend on both the \emph{expertise} of the intended audience, which drives their expectations, and the agent's \emph{competence}, which drives failure to meet reasonable expectations.

However, another possibility is that the audience really did want 
\LimDey{Why} explanations, but the 
casters rarely provided them because of the time they required  --- both theirs and the audience's.
The shoutcasters explained in \emph{real time} as the players performed their actions.
It takes time to understand the present, predict the future, and link present to future; and spending time in these ways reduces the time allowable for explaining interesting activities happening in present.
The corpus showed
casters interrupting themselves and each other as new events transpired, as they tried to keep up with the time constraints. 
This also has implications to the audience's workflow, because it takes time for the audience to mentally process shoutcasters' departures from the present, particularly when interesting actions continuously occur. 

\boldify{Effort too high is based on a compositional argument; Why is COMPOSED of what + what could happen.}

Even more critical to an explanation system, \LimDey{Why} questions also tend to require extra effort (cognitive or computing resources), because they require connecting two time slices: 
\quotateInset{10}{``After seeing the first phoenix and, of course, the second one confirmed, Snute is going to invest in a couple spore crawlers.''}
In this example, the casters had to connect past information (scouting the phoenix, a flying unit) with a prediction of the future (investing in spore crawlers, an air defense structure). 

Answering \LimDey{Why-didn't} questions was even rarer than answering \LimDey{Why} questions (Table~\ref{table:limDeyCoding}).  
Like \LimDey{Why} questions,  \LimDey{Why-didn't} questions required casters to make a connection between previous game state and a potential current or future game state.
For example, \quotate{2}{The probe already left a while ago, so we knew it wasn't going to be a pylon rush.}
\LimDey{Why-didn't} answers' rarity is consistent with the finding that understanding a \LimDey{Why-didn't} explanation requires even more mental effort than a \LimDey{Why} explanation~\cite{lim2009}.
As for an interactive explanation system, supporting \LimDey{Why} questions requires solving both a \emph{temporal credit assignment problem} (determining the effect of an action taken at a particular time on the outcome) and a \emph{structural} one (determining the effect of a particular system element on the outcome).
See~\cite{agogino2004unifying} for an accessible explanation of these problems.



\input{tables/TableContentFreq}

\boldify{Conversely, What is easy to answer, because it has lower cognitive effort AND is also useful, because it has play-by-play and details about game state. So, shoutcasters came 'close enough' by building a Why-ish like answer out of What+WhatCouldHappen. }

The casters found a potentially ``satisficing'' approximation of \LimDey{Why}, a combination of \LimDey{What} and \LimDey{What-could-happen}, 
the two most frequent explanation types.
Their \LimDey{What} answers explained what the player did, what happened in the game, and description of the game state.
These were all things happening in the present,
and did not require the additional cognitive steps required to answer \LimDey{Why} or \LimDey{Why-didn't}, which may have contributed to its high frequency.
Further, the audience needed this kind of ``play-by-play'' information to stay informed about the game's progression; for example, \quotate{4}{This one hero, marine, is starting to kill the vikings}
When adding on \LimDey{What-could-happen}, casters were pairing \LimDey{What} with what the player will or could do, i.e., a hypothetical outcome. For example,
\quotateInset{1}{``...if he gets warning of this he'll be able to get back up behind his wall in.''}
Although answering the question \LimDey{What-could-happen} required predicting the future, it did not also require the casters to tie together information from \emph{past} and future.  

\boldify{``How good/bad was that action'' was able to potentially replace ``why'' because...}

The other two frequent answers, 
\LimDey{How-good/bad-was-that-}\\\LimDey{action}
and \LimDey{How-to,} also sometimes contained ``why'' information. 
For \LimDey{How-good/bad-was-that-action,} casters  \emph{judged} an action e.g.: 
\quotate{1}{Nice maneuver from Jjakji, he knows he can't fight Neeb front on right now, he needs to go around the edges}
For \LimDey{How-to,} casters gave the audience tips and explained high level strategies.
For example, consider this rule-like explanation, which implies the reason ``why'' the player used a particular army composition:\quotate{10}{Roach ravager in general is really good..}

\FIXME{this paragraph has some manual line breaking, ensure it looks ok}

The next rule-like \LimDey{How-to} example is an even closer approximation to ``why'' information. \quotate{8}{Obviously when there are 4 protoss units on the other side of the map, you need to produce more zerglings, which means even fewer drones for Iasonu}
In this case, the casters are giving a rule: given a general game state (protoss units on their side of the map) the player should perform an action (produce zerglings).
But the example does more; it also implies a \LimDey{Why} answer to the question ``Why isn't Iasonu making more drones?''
Since this implied answer simply relates the present to a rule or best practice, it was produced at much lower expense than a true \LimDey{Why} answer that required tying past events to the present.

Mechanisms casters used to circumvent the need for disruptive and resource-intensive \LimDey{Why} explanations, such as using \LimDey{How-to}, may also be ways to alleviate the same problems in explanation systems.

%% file: figures/FigLimDeyFreq.tex
\begin{figure}
	\centering
	\includegraphics[width=\columnwidth]{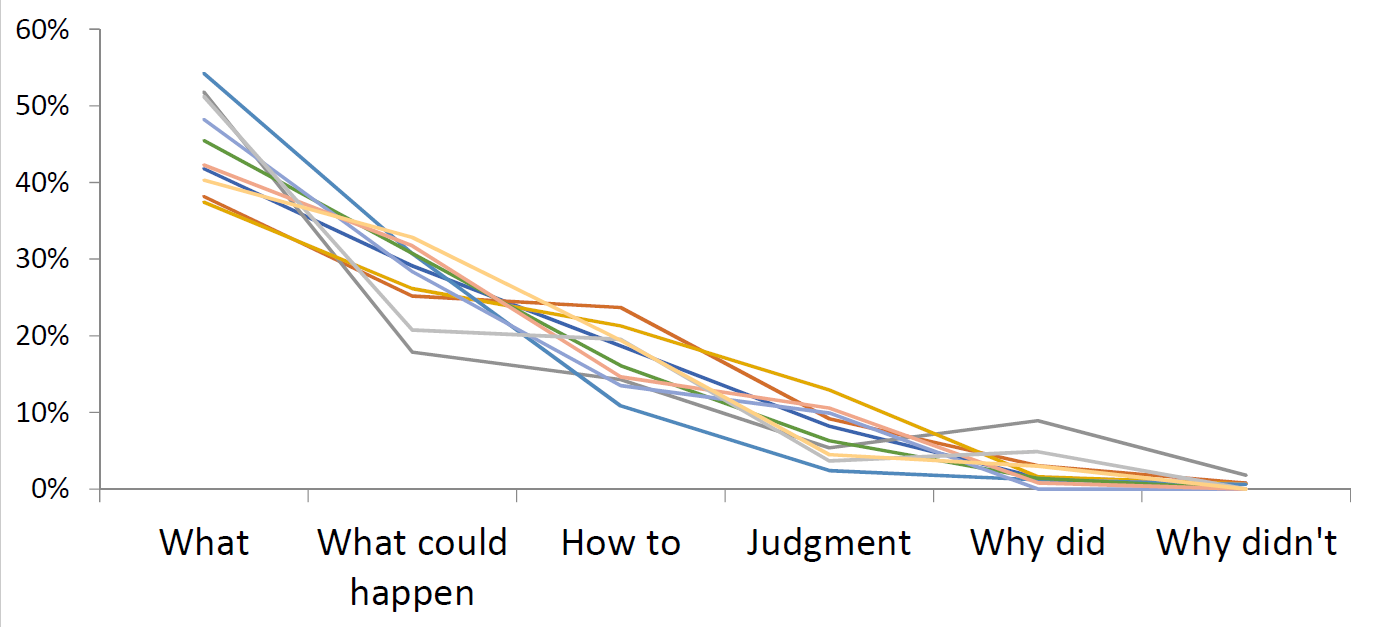}
	\caption{Frequency of Lim \& Dey questions answered by casters, with one line per caster pair.
    Y-Axis represents percentages of the utterances which answered that category of question (X-Axis). 
    Note how casters structured answers consistently.
    }
	\label{fig:limDeyFreq}
\end{figure}

%% file: tables/TableLimDeyCoding.tex
\begin{table*}[t]
	\centering
	\begin{tabular}{@{}p{.145\linewidth}|p{.03\linewidth}|p{.4\linewidth}|p{.355\linewidth}@{}}
		\textit{\textbf{Code}} & 
        \textit{\textbf{Freq}} &
		\textit{\textbf{Description}} &   	
        \textit{\textbf{Example}}
        \\ \hline
		\LimDey{What} & 595
        & What the player did or anything about game state 
        & ``The liberators are moving forward as well'' 
		\\ \hline
		\LimDey{What-could-} & 376
        & What the player could have done or what will happen 
        & ``Going to be chasing those medivacs away'' 
		\\\LimDey{happen} & &
        \\ \hline
		\LimDey{How-to} & 233
        & Explaining rules, directives, audience tips, high level strategies 
        & ``He should definitely try for the counter attack right away'' 
		\\ \hline
		\LimDey{*How-good/bad-} & 112
        & Evaluation of player actions 
        & ``Very good snipe there for Neeb'' 
        \\\LimDey{was-that-action} & &
		\\ \hline
		\LimDey{Why-did} & 27
        & Why the player performed an action 
        & ``...that allowed Dark to hold onto that 4th base, it allowed him to get those ultralisks out'' 
		\\ \hline
		\LimDey{Why-didn't} & 6
        & Why the player did not perform an action 
        & ``The probe already left a while ago, so we knew it wasn't going to be a pylon rush'' 
	\end{tabular}
	\caption{Utterance type code set, slightly modified from the schema proposed by Lim \& Dey.  
    The asterisk denotes the code that we added, \LimDey{How-good/bad-was-that-action} because the casters judged actions based on their quality.}
	\label{table:limDeyCoding}   	
\end{table*}

%% file: tables/TableContentFreq.tex
\begin{table*}
\centering
	\includegraphics[width=.95\textwidth]{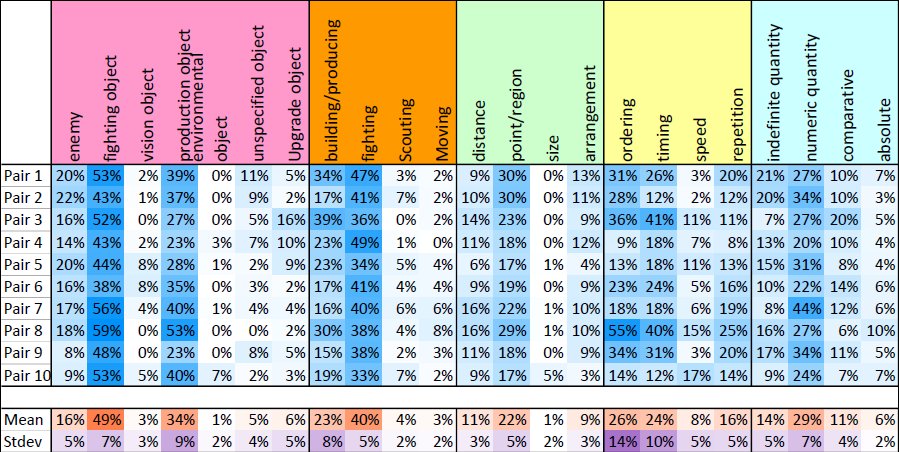}
	\caption{Occurrence frequencies for each code, as a percent of the total number of utterances in the corpus. From left to right: \Ccode{Object} (pink), \Ccode{Action} (orange), \Ccode{Spatial} (green), \Ccode{Temporal} (yellow), and \Ccode{Quantitative} (blue) codes. The casters were consistent about kinds of the content they rarely included, but inconsistent about the kinds of content they most favored.}
	\label{table:contentFreq}
\end{table*}

%% file: 7-ResultsContent.tex
\subsection{RQ4 Results: What relationships and objects do shoutcasters use when building their explanations?}

\boldify{To answer RQ4, we coded, so as to learn what content shoutcasters used to build their explanations, which will inform how sentences should be structured by future explanation systems. Shoutcasters were consistent in their usage of things which were infrequently used and inconsistent in their usage of things which were frequently used.}

To inform future explanation systems' content by expert explanations --- 
the patterns of nouns, verbs, and adjectives/adverbs in these professionally crafted explanations --- we drew upon a code set from prior work~\cite{metoyer2010explaining} (see the Methodology section).
Table~\ref{table:contentFreq} shows how much caster pairs used each of these types of content, grouping the objects (nouns) in the first group of columns, then actions (verbs), and then properties (adjectives and adverbs). 
Table~\ref{table:coOccurrence} shows how the casters' explanations used these concepts \emph{together}, i.e., which properties they paired with which objects and actions.



\boldify{The most important objects were fighting, production, and enemy objects, while the most important actions were fighting and building/producing. This suggests which objects and actions should be described to users by an explanation system.}

The casters' explanation sentences tended to be noun-verb constructions, so we began with the nouns.
The most frequently described objects were \Ccode{fighting object}, \Ccode{production object}, and \Ccode{enemy}, with frequencies of 53\%, 40\%, and 9\%, respectively, as shown in Figure~\ref{table:contentFreq}.
(This is similar to results from \cite{metoyer2010explaining}, where production, fighting, and enemy objects were the three most popular object subcodes.) 
As to the actions (``verbs''), the casters mainly discussed \Ccode{fighting} (40\%) and \Ccode{building} (23\%).
It is not surprising that the casters frequently discussed \Ccode{fighting}, since combat skills are important in StarCraft~\cite{kim2016evaluation}, and \Ccode{producing} is often a prerequisite to \Ccode{fighting}.
This may suggest that, in RTS environments, an explanation system may be able to focus on only the most important subset of actions and objects, without needing to track and reason about most of the others.

\boldify{Co-occurences of object and action subcodes with spatial, temporal, and quantitative subcodes highlight PROPERTIES of our nouns and verbs}

The casters were quite strategic in how they put together these nouns and verbs with properties.
The casters' used particular properties with these nouns and verbs to paint the bigger picture of how the game was going for each player, and how that tied to the players' strategies.  
We illustrate in the next subsections a few of the ways casters communicated about player decisions --- succinctly enough for real time.


\paragraph{``This part of the map is mine!'': Spatial properties}

\boldify{The position of your units define the space you control during combat. Thus, arrangement is similar to point region but at a smaller scale (referenced against units) vs static map elements or structures)}

RTS players claim territory in battles with the \Ccode{arrangement} of their military units, e.g.:
\quotateInset{3}{``He's actually arcing these roaches out in such a great way so that he's going to block anything that's going to try to come back.''}
As the \Ccode{arrangement} column of Table~\ref{table:coOccurrence} shows, the objects that were used most with \Ccode{arrangment} were \Ccode{fighting objects} (12\%, 72 instances) and \Ccode{enemy}, (10\%, 26 instances).
Note that \Ccode{arrangement} is very similar to \Ccode{point/region}, but at a smaller scale;
\Ccode{Arrangement} of \Ccode{production object}, such as exactly where buildings are placed in one's base, appeared to be less significant, co-occurring only 5\% of the time.

\boldify{The players choice of base locations strongly impacts their available strategic choices, thus DISTANCE properties were often modifying [potential] base locations, i.e. REGION}

The degree to which an RTS player wishes to be aggressive or passive is often evident in their choice of what \Ccode{distance} to keep from their opponent, and the casters often took this into account in their explanations.
One example of this was evaluation of potential new base locations.
\quotateInset{5}{``...if he takes the one [base] thats closer that's near his natural [base], then it's close to Innovation so he can harass.''}
Here, the casters communicated the control of parts of the map by describing \emph{bases} as a \Ccode{region}, and then relating two regions with a \Ccode{distance}.
The magnitude of that distance then informed whether the player was able to more easily attack.
Of the casters' utterances that described \Ccode{distance} along with \Ccode{production object}, 27 out of 44 referred to the distance between bases or moving to/from a base.

\paragraph{``When should I...'': Temporal properties}

\boldify{Speed illuminated the player's intended strategy by showing their priorities while BUILDING, e.g., Expanding early means the player intends to focus on economy instead of being aggressive during the early/mid game. }

Casters' explanations often reflected players' priorities for allocating limited resources.
One way they did so was using \Ccode{speed} properties:
\quotate{4}{We see a really quick third [base] here from XY, like five minutes third}
Since extra bases provide additional resource gathering capacity, the audience could infer that the player intended to follow an ``economic'' strategy, as those resources could have otherwise been spent on military units or upgrades.
This contrasts with the following example, 
\quotate{8}{He's going for very fast lurker den..}
The second example indicated the player's intent to follow a different strategy: unlocking stronger units (lurkers).
\Ccode{Speed} co-occurred with \Ccode{building/producing} most often (12\%, 36 instances).



\input{tables/TableCoOccurrence}

\paragraph{``Do I care how many?'': Quantitative properties}

\boldify{Surprise! At the lowest level of specificity (comparative), shoutcasters tend to avoid mentioning the TYPE of objects.  Much of this is due to comparative discussion of ``supply'' or ``units'' (generic terms to aggregate their holdings)}

We found it surprising how often the casters described quantities without numbers.
In fact, the casters often did not even include \emph{type} information when they described the players' holdings, instead focusing on \Ccode{comparative} properties (Table~\ref{table:coOccurrence}).
For example, \quotate{1}{There is too much supply for him to handle. Neeb finalizes the score here after a fantastic game} 
Here, ``supply'' is so generic, we do not even know what kind of things Neeb had -- only that he had ``too much'' of it.

\boldify{The next level of specificity: Cheap military units are given with TYPE information, but often VAGUE numbers.  Possibly because they are usually numerous, and the exact count is hard to 1. derive or 2. access.}

In contrast, when the casters discussed cheap military units, like ``marines'' and ``zerglings,'' they tended to provide \emph{type} information, but about half of their mentions still included no precise numbers.
Perhaps it was a matter of the high cost to get that information: cheap units are often built in large quantities, so deriving a precise quantity is often very tedious.
Further, adding one weak unit that is cheap to build has little impact on army strength, so getting a precise number may not have been worthwhile -- i.e. the value of knowing precise quantities is low. 
To illustrate, consider the following example, which quantified the army size of both players vaguely, using \emph{indefinite quantity} properties:
\quotate{6}{That's a lot of marines and marauders and not enough stalkers}

\boldify{The highest level of specificity: workers are given TYPE information and PRECISE numbers.  Possibly because the exact count is easily accessible. Further, workers are extremely important.}

In the RTS domain, workers are a very important unit.
Consistent with this importance, workers are the only unit where the casters were automatically alerted to their death (Figure~\ref{fig:gameScreenshot}, region 4), and are also available at a glance on the HUD (Figure~\ref{fig:gameScreenshot} region 1).
Correspondingly, the casters often gave precise quantities of workers (a \Ccode{production object}).
Workers (workers, drones, scvs, and probes) had 46 co-occurrences with \Ccode{numeric quantities}, but only 12 with \Ccode{indefinite quantities}  (e.g., lot, some, few).
\quotate{2}{...it really feels like Harstem is doing everything right, and [yet] somehow ended up losing 5 workers}


\subsubsection{\ImplicationsText}

\boldify{Meta boldify: we have essentially 4 take aways: 
\\1) content coding can be used to build sentences by considering favorite concepts and concept co-ocurrence [but we beat this to death already] 
\\2) spatial properties both help describe property ownership, but also reveal domain strategy because distance can indicate intended plans to attack or defend 
\\3) Speed also reveals domain strategy in terms of resource allocation 
\\4) Quantity helps us choose the right level of abstraction to use for a group of units}

These results have particularly important implications for interactive explanation systems with real-time constraints.
Namely, the results suggest that an effective way to communicate about strategies and tactics is to modify the critical objects and actions with particular properties that suggest strategies.  
This not only affords a \emph{succinct} way to communicate about strategies and tactics (fewer words) but also a \emph{lighter load} for both the system and the audience than attempting to build and process a rigorous explanation of strategy.

Specifically, spatial properties can communicate beyond the actual properties of objects to strategies themselves; for example, casters used distance to point out plans to attack or defend.
Temporal properties can be used in explanations of strategies when choices in resource allocation determines available strategies. 

Finally, an interactive explanation system could use the quantitive property results to help ensure alignment in the level of abstraction used by the human and the system.
For example, a player can abstract a quantity of units into a \emph{single group} or think of them as \emph{individual units}.
Knowing the level of abstraction the human players use in different situations can help an interactive explanation system choose the level of abstraction that will meet human expectations.
Using properties in these strategic ways may enable an interactive explanation system to meet its real-time constraints while at the same time improving its communicativeness to the audience.


%% file: tables/TableCoOccurrence.tex
\begin{table}
\centering
	\includegraphics[width=\linewidth]{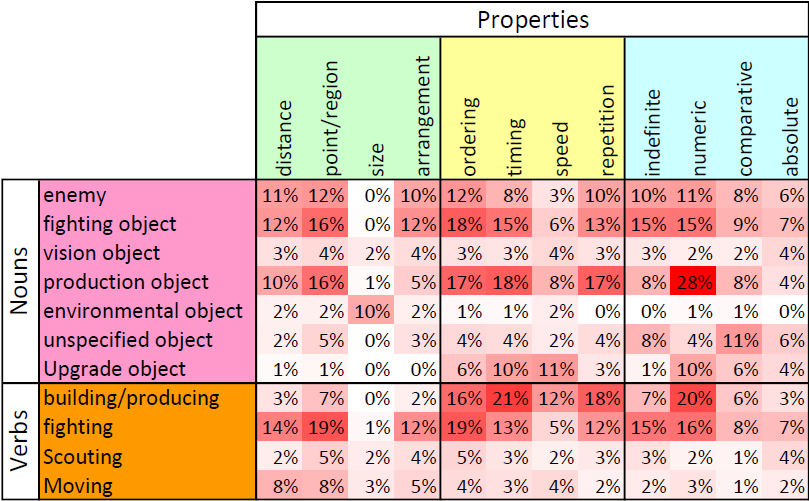}
	\caption{Co-Occurrence Matrix. 
    Across rows: \Ccode{Object} (pink, top rows) and \Ccode{Action} (orange, bottom rows) codes.
    Across columns: \Ccode{Spatial} (green, left), \Ccode{Temporal} (yellow, center), and \Ccode{Quantitative} (blue, right). 
    Co-occurrence rates were calculated by dividing the intersection of the subcodes by the union. }
	\label{table:coOccurrence}
\end{table}

%% file: 8-Conclusion.tex
\section{Conclusion}


The results of our study suggest that explaining intelligent agents to humans has much to gain from looking to the human experts.   
In our case, the expert explainers --- RTS shoutcasters --- revealed implications into  what, when, and how human audiences of such systems need explanations, and how real-time constraints can come together with explanation-building strategies.    
Among the results we learned were:

\vspace{-5pt}
\begin{enumerate}[labelindent=20pt,labelwidth=\widthof{\ref{last-item-2}},label=\arabic*.,itemindent=1em,leftmargin=!]
\item[\textbf{RQ1}] 
Investigating the what's and where's of casters' real-time information foraging to \emph{assess and understand} the players showed that the most commonly used \emph{patches} of the information environment were the Actuators (``A'' in the PEAS model). This suggests that, in contrast to today's explanation systems, which tend to present mostly Performance measures, explanations should consider presenting more from the Actuators and Sensors.
\vspace{-5pt}
\item[\textbf{RQ2}] 
The how's of casters' foraging revealed a common pattern, which we termed the A-E-P+S loop, and the most common cues and triggers that led shoutcasters to move through this loop. Future explanation systems may be well-served to prioritize and recommend explanations according to this loop and its triggers.
\vspace{-5pt}
\item[\textbf{RQ3}] 
As \emph{model explainer}, the casters revealed strategies for ``satisficing'' with explanations that may not have precisely answered all the questions the audience had in mind, but were feasible given the time and resource constraints in effect when comprehending, assessing, and explaining, all in \emph{real time} as play progresses. These strategies may be likewise applicable to interactive explanation systems.
\vspace{-5pt}
\item[\textbf{RQ4}] 
The detailed contents of the casters' explanations revealed  patterns of how they paired properties (``adjectives and adverbs'') with different objects (``nouns'') and actions (``verbs''). Interactive explanation systems may be able to leverage these patterns to communicate succinctly about an agent's tactics and strategies. 
\label{last-item-2}
\end{enumerate}
\vspace{-5pt}

Ultimately, both shoutcasters' and explanation systems' jobs are to improve the audience members' mental model of the agents' behavior.
As Cheung, et al.~\cite{Cheung:2011:SSU:1978942.1979053} put it, ``...commentators are valued for their ability to expose the depth of the game.''
Hopefully, future explanation systems will be valued for the same reason.



\section{Acknowledgments}
Removed for anonymous review.